\documentclass[a4paper, amsfonts, amssymb, amsmath, reprint, showkeys, floatfix, nofootinbib, twoside, superscriptaddress]{revtex4-2}
\usepackage[english]{babel}
\usepackage[utf8]{inputenc}
\usepackage[colorinlistoftodos, color=green!40, prependcaption]{todonotes}
\usepackage{xcolor}
\usepackage{siunitx}
\usepackage{textcomp}
\usepackage{gensymb}
\usepackage{array}
\usepackage{longtable}
\usepackage{booktabs}
\usepackage{amsthm}
\usepackage{mathtools}
\usepackage{physics}
\usepackage{xcolor}
\usepackage{graphicx}
\usepackage[left=23mm,right=13mm,top=35mm,columnsep=15pt]{geometry} 
\usepackage{adjustbox}
\usepackage{placeins}
\usepackage[T1]{fontenc}
\usepackage{lipsum}
\usepackage{csquotes}

\AtBeginDocument{\RenewCommandCopy\qty\SI} 
\usepackage[pdftex, pdftitle={Article}, pdfauthor={Author}]{hyperref} 
\newcommand\Tstrut{\rule{0pt}{2.6ex}}         
\newcommand\Bstrut{\rule[-0.9ex]{0pt}{0pt}}   

\AtBeginDocument{\RenewCommandCopy\qty\SI} 
\begin{document}

\title{Emergent pair localization in a many-body quantum spin system}

\author{Titus Franz}
\email[]{franzt@physi.uni-heidelberg.de}
\affiliation{Physikalisches Institut, Universit\"at Heidelberg, Im Neuenheimer Feld 226, 69120 Heidelberg, Germany}
\affiliation{Max-Planck-Institut für Quantenoptik, Hans-Kopfermann-Str. 1, Garching, Germany}
\author{Sebastian Geier}
\affiliation{Physikalisches Institut, Universit\"at Heidelberg, Im Neuenheimer Feld 226, 69120 Heidelberg, Germany}
\author{Adrian Braemer}
\affiliation{Physikalisches Institut, Universit\"at Heidelberg, Im Neuenheimer Feld 226, 69120 Heidelberg, Germany}
\affiliation{Kirchhoff-Institut für Physik, Universität Heidelberg, Im Neuenheimer Feld 227, 69120 Heidelberg, Germany}
\author{Cl\'ement Hainaut}
\affiliation{Physikalisches Institut, Universit\"at Heidelberg, Im Neuenheimer Feld 226, 69120 Heidelberg, Germany}
\affiliation{Univ. de Lille, CNRS, UMR 8523–PhLAM–Physique des Lasers, Atomes et Molécules, Lille, France}
\author{Adrien Signoles}
\affiliation{Physikalisches Institut, Universit\"at Heidelberg, Im Neuenheimer Feld 226, 69120 Heidelberg, Germany}
\affiliation{PASQAL, 7 rue Léonard de Vinci, 91300 Massy, France}
\author{Nithiwadee Thaicharoen}
\affiliation{Physikalisches Institut, Universit\"at Heidelberg, Im Neuenheimer Feld 226, 69120 Heidelberg, Germany}
\affiliation{Department of Physics and Materials Science, Faculty of Science, Chiang Mai University, 239 Huay Kaew Road, Muang, Chiang Mai, 50200, Thailand}
\author{Annika Tebben}
\affiliation{Physikalisches Institut, Universit\"at Heidelberg, Im Neuenheimer Feld 226, 69120 Heidelberg, Germany}
\author{Andre Salzinger}
\affiliation{Physikalisches Institut, Universit\"at Heidelberg, Im Neuenheimer Feld 226, 69120 Heidelberg, Germany}
\author{Martin Gärttner}
\email[]{martin.gaerttner@uni-jena.de}
\affiliation{Physikalisches Institut, Universit\"at Heidelberg, Im Neuenheimer Feld 226, 69120 Heidelberg, Germany}
\affiliation{Kirchhoff-Institut für Physik, Universität Heidelberg, Im Neuenheimer Feld 227, 69120 Heidelberg, Germany}
\affiliation{Institut für Theoretische Physik, Ruprecht-Karls-Universität Heidelberg, Philosophenweg 16, 69120 Heidelberg, Germany}
\affiliation{Institute of Condensed Matter Theory and Optics, Friedrich-Schiller-University Jena, Max-Wien-Platz 1, 07743 Jena, Germany}
\author{Gerhard Zürn}
\affiliation{Physikalisches Institut, Universit\"at Heidelberg, Im Neuenheimer Feld 226, 69120 Heidelberg, Germany}
\author{Matthias Weidemüller}
\email[]{weidemueller@uni-heidelberg.de}
\affiliation{Physikalisches Institut, Universit\"at Heidelberg, Im Neuenheimer Feld 226, 69120 Heidelberg, Germany}

\date{\today} 

\begin{abstract}

Understanding how closed quantum systems dynamically approach thermal equilibrium presents a major unresolved problem in statistical physics. 
Generically, non-integrable quantum systems are expected to thermalize as they comply with the Eigenstate Thermalization Hypothesis. However, in the presence of strong disorder, the dynamics can possibly slow down to a degree that systems fail to thermalize on experimentally accessible timescales, as in spin glasses or many-body localized systems. 
In general, particularly in long-range interacting quantum systems, the specific nature of the disorder necessary for the emergence of a prethermal, metastable state—distinctly separating the timescales of initial relaxation and subsequent slow thermalization—remains an open question.  
We study an ensemble of Heisenberg spins with a tunable distribution of random coupling strengths realized by a Rydberg quantum simulator. 
We observe a drastic change in the late-time magnetization when increasing disorder strength. The data is well described by models based on pairs of strongly interacting spins, which are treated as thermal for weak disorder and isolated for strong disorder. Our results indicate a crossover into a pair-localized prethermal regime in a closed quantum system of thousands of spins in the critical case where the exponent of the power law interaction matches the spatial dimension. 
\end{abstract}

\maketitle

\section{Introduction}

\begin{figure*}[!htbp]
    \centering
    \includegraphics[scale=1]{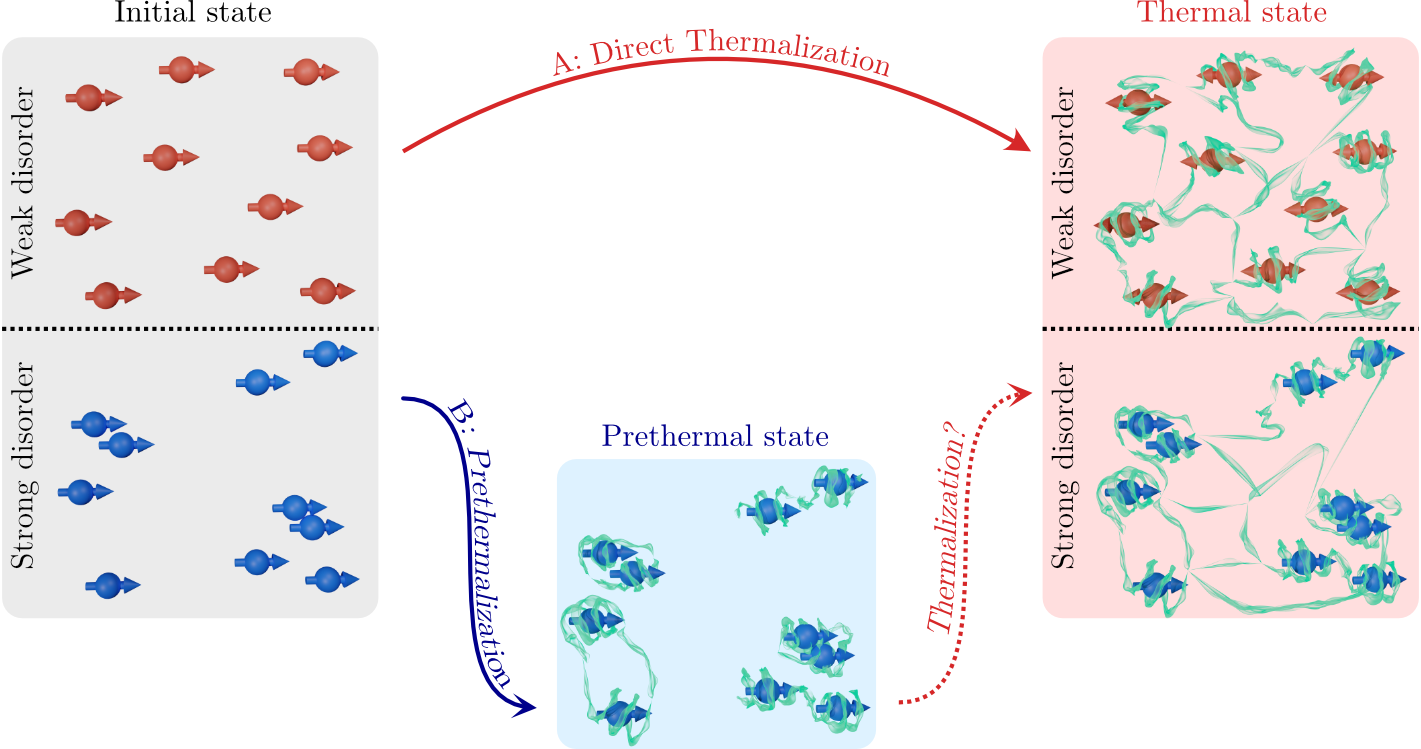}
    \caption{\textbf{Thermal and non-thermal regimes.} Schematic depiction of the dynamics of the system. Depending on the strength of disorder, initially, uncorrelated spins evolve either directly (via \textbf{A}) to a thermal state with correlations between all spins or (via \textbf{B}) to a prethermal state consisting of uncorrelated pairs of strongly correlated spins. Whether, in this case, thermalization occurs eventually remains an open question.}
    \label{fig:figure1}
\end{figure*}

What is the fate of an isolated, strongly interacting, and possibly disordered quantum system initially prepared in a far-from-equilibrium state? In general, even if a strongly interacting quantum system is isolated from its environment, it is expected to thermalize~\cite{reimannFoundationStatisticalMechanics2008, deutsch1991, srednicki, srednickiChaosQuantumThermalization1994}. 
As a notable exception to this rule, strongly disordered systems can retain retrievable memory of their initial state for arbitrarily long times, leading to a rich phenomenology ranging from glassy dynamics to many-body localization.

A comprehensive framework for understanding systems that do not undergo direct thermalization is provided by the concept of prethermalization~\cite{Ueda2020QuantumEquilibration, peng_floquet_2021, prufer2018, rubio-abadal_floquet_2020, eigen2018, martinControllingLocalThermalization2022}: Here, the Hamiltonian can be decomposed into a reference Hamiltonian $H_0$, and a weak perturbation $H_1$ which breaks at least one local conservation law of $H_0$. In such instances, a metastable state exists whose properties can be calculated using the generalized Gibbs ensemble (GGE) of the reference Hamiltonian $H_0$~\cite{langenPrethermalizationUniversalDynamics2016}. As an example, we can consider many-body localization (MBL) in the framework of prethermalization: Here, the reference Hamiltonian is given by a non-interacting ensemble of spins subject to a strongly disordered external field, and $H_1$ describes the interactions between nearest-neighbor spins. If these interactions are sufficiently weak, these systems remain localized~\cite{baskoMetalInsulatorTransition2006, Imbrie2016}, and the conserved quantities become “dressed,” commonly referred to as l-bits~\cite{Serbyn2013, Huse2014}. However, a different type of disorder naturally occurs in numerous systems, including cold atoms~\cite{kondovDisorderInducedLocalizationStrongly2015, schreiberObservationManybodyLocalization2015, bordiaCouplingIdenticalOnedimensional2016, choiExploringManybodyLocalization2016, kaufmanQuantumThermalizationEntanglement2016, lukinProbingEntanglementManybody2019, wadleighInteractingStarkLocalization2022, leonardSignaturesBathinducedQuantum2022}, ions~\cite{smithManybodyLocalizationQuantum2016} or nitrogen-vacancy centers~\cite{kucskoCriticalThermalizationDisordered2018, martinControllingLocalThermalization2022} where the couplings themselves are disordered, not the external field. In this case, discerning the reference Hamiltonian $H_0$ becomes nontrivial, and in previous studies, the depolarization dynamics in these systems is interpreted as direct thermalization~\cite{kucskoCriticalThermalizationDisordered2018, martinControllingLocalThermalization2022}. Yet, it is known from spin glasses that disorder in the couplings leads to a hierarchy of timescales, which slows down the dynamics such that on experimentally accessible timescales, these systems never reach thermal equilibrium. 

Unfortunately, understanding if and how quantum systems thermalize is extremely challenging, as numerical simulations are limited to relatively small system sizes~\cite{suntajsErgodicityBreakingTransition2020, suntajsQuantumChaosChallenges2020, sierantChallengesObservationManybody2022} and analytical solutions are scarce. Quantum simulation experiments with single-site resolution can investigate slightly larger systems with several tens of particles~\cite{Bloch2008ManybodyPhysicsa, Esslinger2010FermiHubbardPhysics, Bloch2012QuantumSimulations, Gross2017QuantumSimulationsa, Browaeys2020ManybodyPhysics}, but they can only probe finite time scales~\cite{schreiberObservationManybodyLocalization2015, bordiaCouplingIdenticalOnedimensional2016, choiExploringManybodyLocalization2016, smithManybodyLocalizationQuantum2016}.

In this study, we employ a Rydberg quantum simulator~\cite{browaeysManybodyPhysicsIndividually2020, signolesGlassyDynamicsDisordered2021, geierFloquetHamiltonianEngineering2021} to explore the thermalization dynamics in long-range interacting systems in 3D using a cloud of up to 6000 Rydberg spins with spatial disorder. 
In the weakly disordered regime, characterized by similar distances between particles (top row in Fig. 1), our experimental findings align with previous assertions of direct thermalization~\cite{kucskoCriticalThermalizationDisordered2018}. However, at strong disorder, we demonstrate the emergence of a localized prethermal state, verified through a non-analytical dependence of the late-time magnetization on an external field. Here, the hierarchy of interaction strengths allows us to effectively describe the Hamiltonian with a reference Hamiltonian $H_0$, where strongly interacting pairs of spins remain localized for long times before interactions between pairs possibly lead to thermalization at even later times (bottom row in Fig. 1).

\section{Experimental setup}

We consider the quantum spin-1/2 Heisenberg XXZ-model (in units where $\hbar=1$) 
\begin{equation}
	\hat H_{\text{int}} = \frac{1}{2} \sum_{i, j} J_{ij}\left( \hat{s}_x^{(i)}\hat{s}_x^{(j)} + \hat{s}_y^{(i)} \hat{s}_y^{(j)} + \delta \hat{s}_z^{(i)} \hat{s}_z^{(j)}  \right) \, ,
	\label{eq:H_int}
\end{equation}
with spin operators $\hat{s}_{\alpha}^{(i)}=\hat{\sigma}_{\alpha}^{(i)}/2$ ($\alpha \in \{ x, y, z\}$) acting on spin $i$.
The interactions between spins decay with a power law $J_{ij} = C_{a}r_{ij}^{-a}\left( 1 - 3\cos(\theta_{ij})^2 \right)$, where $r_{ij}$ are the distances between the spins $i$ and $j$ and $\theta_{ij}$ is the angle with respect to the quantization axis defined by the magnetic field. The parameters $\delta$ and $a$ are determined by the choice of Rydberg states (cf.~\cite{signolesGlassyDynamicsDisordered2021}). In our experiment, we encode the spin degree of freedom in the Rydberg states $\ket{\downarrow} = \ket{48S}$ and $\ket{\uparrow} = \ket{48P}$ leading to dipolar interactions as described by Eq.~\eqref{eq:H_int} with  $\delta = \num{0}$, $a=3$ and $C_3/(2\pi) = \SI{1.15}{\giga\hertz \micro\meter\tothe{3}}$. Additional data for a Van-der-Waals interacting system ($a=6$, $C_6/(2\pi) = \SI{507}{\giga\hertz \micro\meter\tothe{6}}$, $\delta\approx -0.7$ and no angular dependence on $\theta_{ij}$) is shown in Appendix \ref{appendix:vdW-data}.

The spins are distributed randomly with an imposed minimal distance $r_{\mathrm{bl}}$ resulting in a random but correlated distribution of couplings $J_{ij}$ (Fig.~\ref{fig:figure1}). This geometry is naturally given in the experiment where the Rydberg blockade effect forbids two excitations being closer than $r_{\mathrm{bl}}$. The blockade constraint allows tuning the strength of the disorder: 
For the weak disorder measurements, we chose the density such that the typical interparticle distance of $a_0 \sim \SI{6.8}{\micro \meter}$ is comparable to the blockade radius of \SI{4.6}{\micro\meter}, whereas in the strongly disordered case, the blockade radius of \SI{5.0}{\micro\meter} is much smaller compared to the typical interparticle distance $a_0 \approx \SI{11.2}{\micro \meter}$ (see methods for more details on the Rydberg atoms' distribution). In both cases, the median interaction strengths of $J_{\mathrm{median}}/(2\pi) = \mathrm{median}_i\left(  \max_j |J_{ij}| \right) / (2\pi) = \SI{2.8}{\mega\hertz}$ (weak disorder) or $J_{\mathrm{median}}/(2\pi) = \SI{1.1}{\mega\hertz}$ (strong disorder) is large compared to typical decoherence rates like the decay rate of the Rydberg atoms of $\Gamma/(2\pi) = \SI{0.018}{\mega\hertz}$. The maximal duration of the experiment of \SI{10}{\micro\second} is chosen such that the Rydberg decay can still be considered small. 

By coupling the spin states with a microwave field $\Omega$, we perform a Ramsey protocol (schematically depicted in Fig.~\ref{fig:figure2}~\textbf{a}) where a first $\pi/2$ pulse initially prepares the system in the fully $x$-polarized state $\ket{\psi_0}=\ket{\rightarrow}_x^{\otimes N} =2^{-N/2}(\ket{\uparrow} +\ket{\downarrow})^{\otimes N}$ (see methods for details of the experimental protocol) which shows no classical dephasing or dynamics in a mean-field description where for each atom, the effective field is aligned with the polarization of the atoms (see Fig.~\ref{fig:figure2}~\textbf{a} center). With a second $\pi/2$-pulse, we read out the average magnetization $\langle \hat{S}_x\rangle = \langle \sum_i\hat{s}_x^{(i)}\rangle/N$. Since this observable is an average over local (single-spin) observables, it should relax to its thermal value if the system is locally thermalizing.

\section{Hierarchy of relaxation time scales}

The blue dots in Fig.~\ref{fig:figure2}~\textbf{b} (strong disorder) and the red dots in Fig.~\ref{fig:figure2}~\textbf{c} (weak disorder) labeled with \SI{0.0}{\mega\hertz} show the time evolution of the magnetization under $\hat H_{\text{int}}$. In both regimes, the magnetization relaxes to zero, following a stretched exponential law as discussed in previous work~\cite{signoles2021, schultzen2021GlassyQuantum, schultzen2021SemiclassicalSimulations}, and reaches a steady-state on a time scale of $\sim 2\pi/J_{\rm median}$ in units of the inverse median nearest neighbor interaction strength. This depolarization dynamics is a direct consequence of the symmetry of the interaction Hamiltonian as all eigenstates already have vanishing $x$-magnetization due to the conservation of $\sum_i \hat{s}_z^{(i)}$\footnote{From the conversation of $\hat{S}_z$, i.e. $[\hat{S}_z, \hat{H}_{\text{int}}]=0$, it follows that $\langle \hat{S}_x\rangle = -i\langle [\hat{S}_y, \hat{S}_z]\rangle = 0$ for every eigenstate.}. 

This situation changes when adding a homogeneous transverse field term to the Hamiltonian
\begin{equation}
	\hat{H}_{\text{ext}} = \Omega \sum_i \hat{s}_x^{(i)},
	\label{eq:H_ext}
\end{equation}
which breaks the $U(1)$ symmetry and leads to a finite late-time magnetization as the data in Fig.~\ref{fig:figure2}~\textbf{b} and \textbf{c} shows. As a result, the dynamics still feature an initial fast relaxation on the time-scale of $2\pi/J_{\mathrm{median}}$, followed by a slowly relaxing regime. 
The stronger the applied magnetic field, the sooner the metastable regime is reached, and the higher the magnetization value becomes.

The finite late-time value of the magnetization of these curves may be understood on a qualitative level by a simple, intuitive, spin-locking model~\cite{Gunther2013NMRSpectroscopy}. At strong field $\Omega\gg J_{\mathrm{median}}$, the inter-spin interaction cannot overcome the magnetic forces and so the spins stay put. Lowering the external field strength weakens this lock and the spins can start dephasing due to their interactions. 

As a consistency check, we compare the experimental data to semiclassical truncated Wigner approximation (dTWA) (solid lines in Fig.~\ref{fig:figure2}~\textbf{b} and \textbf{c}). All simulated curves agree well with the experimental data except for the strongest magnetic field strength, confirming the quality of our quantum simulation of the Heisenberg model. The deviations at strong magnetic field (see grey dots in Fig.~\ref{fig:figure2}~\textbf{b} are likely caused by an experimental imperfection as the strong field may lead to additional couplings to other Rydberg states. This induces population loss to states, which are not measured, thus reducing the total magnetization. Therefore, all experimental data at external field strength above \SI{5}{\mega\hertz} are greyed out.

\section{Prethermalization in disordered spin systems}

\begin{figure*}[!htbp]
    \centering
        \includegraphics[width=\textwidth]{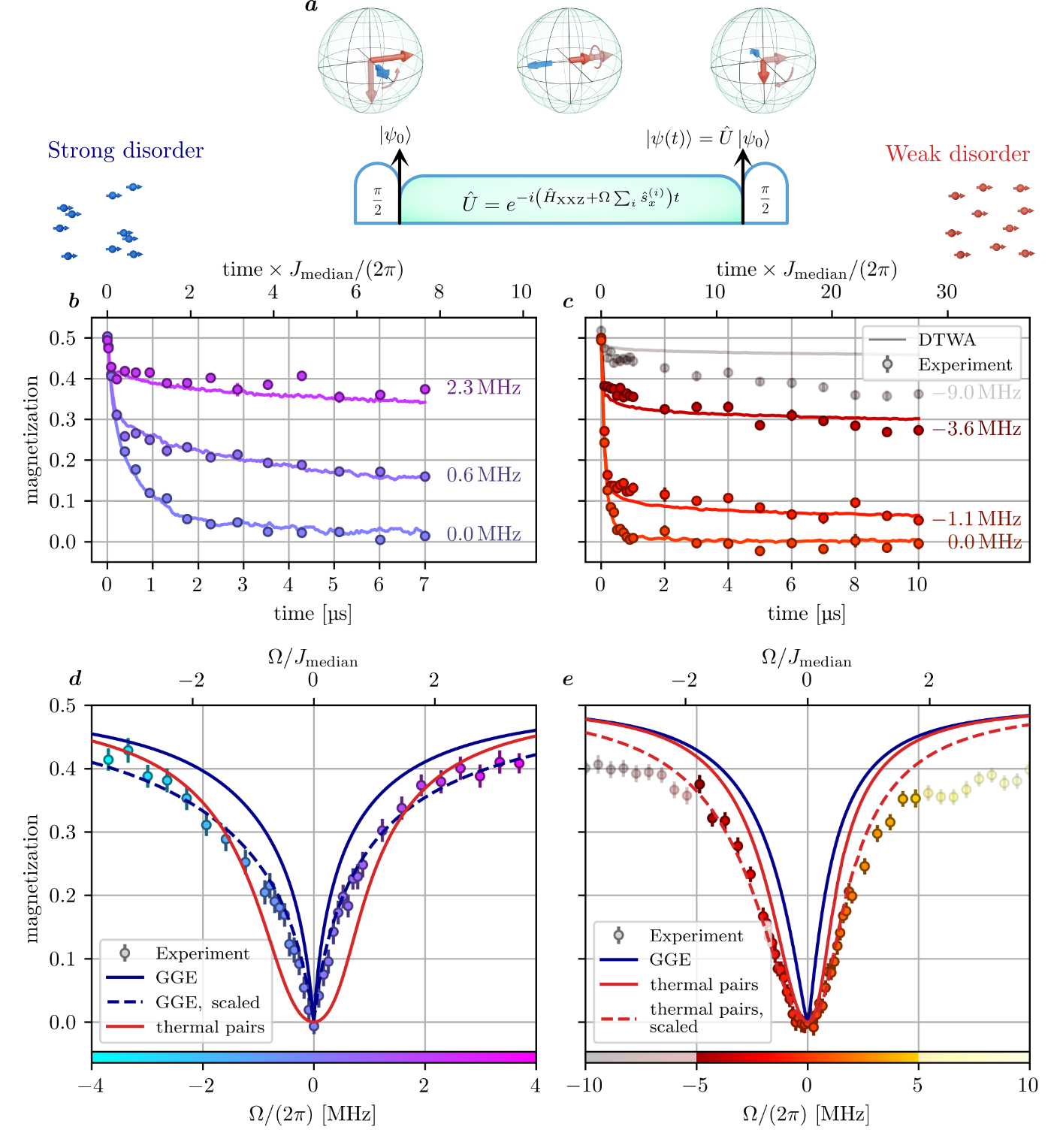}
    \caption{\textbf{Late-time magnetization for different strength of disorder for a spin system interacting with dipole-dipole interactions}. 
    \textbf{a} Experimental protocol: A $\pi/2$ pulse (blue arrow in the Bloch sphere) rotates the spins from the $z$ (light red arrow) to the $x$-direction (red arrow). During the subsequent time evolution, the system interacts via the Heisenberg Hamiltonian~\eqref{eq:H_int} while a spin locking field at Rabi frequency $\Omega$ is applied. The final magnetization is read out after a second $\pi/2$-pulse. 
    \textbf{b} (strong disorder) and \textbf{c} (weak disorder): Measured spin relaxation dynamics for varying transverse field strengths. The solid lines show semiclassical DTWA simulations. \textbf{d} (strong disorder) and \textbf{e} (weak disorder): Magnetization after \SI{10}{\micro\second} as a function of field strength $\Omega$. The solid blue (red) lines show the magnetization expected from a GGE~\eqref{eq:GGE_pair} (canonical ensemble~\eqref{eq:canonical_pair}). The dashed lines show the same simulations rescaled by a global factor to best fit the experimental data.
    }
    \label{fig:figure2}
\end{figure*}

The striking difference between strongly and weakly disordered cases becomes apparent when examining the dependence of the plateau value on the external field measured after \SI{10}{\micro\second} (see Fig.~\ref{fig:figure2}\textbf{d} and \textbf{e}). For strong disorder, there is a sharp cusp at $\Omega=\SI{0}{\mega\hertz}$, which is not present for weak disorder, where the curve is very smooth. Note that this is not an artifact of the difference in absolute scale of the x-axis caused solely by the on average weaker interactions in the strongly disordered case. Relative to their respective median interaction strength $J_{\mathrm{median}}$, both plots cover a similar range. For a generic, thermalizing system, it is plausible to expect a smooth parametric dependence based on the Eigenstate Thermalization Hypothesis (ETH). We will argue that the cusp feature is a clear signature of a non-thermal state, consistent with a generalized Gibbs ensemble with extensively many conserved quantities.

To explain the curve in the strongly disordered regime, characterized by a small blockade radius $r_{\mathrm{bl}}$ and significant variations in the nearest neighbor inter-spin distances, we employ a model based on pairs of strongly interacting spins: For strong positional disorder, close-by spins form pairs that approximately decouple from the rest of the system as the energy splitting between their eigenstates will typically be much larger than any other terms in the Hamiltonian affecting the pair~\cite{Vosk2013ManyBodyLocalization, Pekker2014HilbertGlassTransition, Vasseur2015QuantumCriticality, Vasseur2016ParticleholeSymmetry, braemerPairLocalizationDipolar2022}. In the presence of an external field, such a pair, initially in a fully polarized state, will undergo coherent oscillations and depolarize only very slowly as it does not become entangled with the rest of the system until very late times~\cite{schultzen2021GlassyQuantum}. Thus, the depolarization dynamics arises due to dephasing among pairs oscillating at different frequencies due to different interaction strengths. 

Thus, returning to the notion of prethermalization introduced above, $H_0$ is given by the part of the full Hamiltonian that acts on single pairs, while interactions between the pairs form the weak part $H_1$. Since $H_0$ factorizes into contributions of the individual pairs, we can make a prediction for the prethermal steady state magnetization. 
On average, each pair contributes
\begin{equation}
    \overline{\langle \hat{S}_x \rangle_{\mathrm{pair}}} = \frac{\Omega^2}{2(\Omega^2 + j^2)}
\end{equation}
to the total magnetization, where $j = J\left(\delta - 1\right)$ and $J$ the pair's coupling (see Appendix~\ref{app:pair_model} for a detailed derivation). The distribution of pair couplings can be found numerically by sampling blockaded positions. To calculate the steady-state value, we use a self-consistent mean-field approach to assign each pair an effective field strength $\Omega_i$ taking into account the interaction with its surroundings. This leads to the  asymmetry around $\Omega=0$, as for $\Omega > 0$, the mean-field contribution adds to the static part and thus results in a larger effective field, while for $\Omega < 0$ the converse is true. This effect is much more prominent in the case of $\alpha=6$ as shown in the appendix in Fig.~\ref{fig:van-der-waals}.

In essence, this mean-field pair model describes the system as a generalized Gibbs ensemble
\begin{equation}
\rho_\mathrm{GGE} \propto \exp\left(-\sum_i \beta_i H_\mathrm{pair,i}^{(\text{mf})}\right)
\label{eq:GGE_pair}
\end{equation}
of pairs governed by $H_\mathrm{pair,i}^{(\text{mf})}$, where the Lagrange multipliers $\beta_i$ are fixed by energy conservation. 
Using this model, we find qualitative agreement with the experimental data in the case of strong disorder (blue, solid line in Fig.~\ref{fig:figure2}\textbf{d}). If the interaction strength of the pair simulation is artificially increased by a factor of $1.75$ (dashed blue line), we find even perfect agreement with the experiment. We conjecture that this factor is needed to take into account interactions beyond the nearest neighbor. 

Thus, we have shown that the system is consistent with a prethermal description in the sense, that we found a quasi-stationary state inconsistent with a thermal ensemble description yet matching a generalized Gibbs ensemble. Furthermore, this prethermal state is localized as the pairs' eigenspaces constitute local integrals of motion.

In the less disordered regime, this model of isolated pairs also predicts a sharp, narrow shape (see blue, solid line in Fig.~\ref{fig:figure2}~\textbf{e}) which does not match the experimental data even on a qualitative level. In this regime, the approximation of isolated pairs of spins is no longer valid, and we need to consider the build-up of entanglement between different pairs of spins, which leads to fast thermalization. While the full treatment of the highly correlated many-body system of 6000 spins is not feasible on a classical computer, we can make the first order approximation that the system itself acts as a thermal bath for each pair and imposes that all pairs thermalize to the same global temperature (see also Appendix~\ref{app:pair_model}):
\begin{equation}
\rho_\mathrm{can} \propto \exp\left(-\beta \sum_i  H_\mathrm{pair,i}^{(\text{mf})}\right)
\label{eq:canonical_pair}
\end{equation}
Here, $\beta$ is defined implicitly by energy conservation $\Tr \rho_\mathrm{can} \hat{H} = \expval{\hat{H}}{\psi_0}$. We find qualitative agreement between this model (red, solid line) and experimental data. The agreement can be improved by increasing the interactions by a factor of 1.4 (red, dashed line) that effectively takes into account the correlations between distant spins that are neglected in the pair description of eq.~\ref{eq:canonical_pair}. The deviation at strong field is likely caused by coupling to different Rydberg states as remarked earlier.

As a consistency check, we also try to explain the data in the strong-disorder regime with the canonical ensemble description (see Fig.~\ref{fig:figure2}~\textbf{d}) which clearly fails to reproduce the observed sharp cusp around $\Omega=\SI{0}{\mega\hertz}$.

\section{Conclusion and Discussion}
We studied the relaxation dynamics of power-law interacting spins by observing the change in the parametric dependence of the late-time magnetization on an external field. By finding simple models based on pairs of strongly interacting spins, we explained the measured data both in the weak and strong disorder regime revealing a fundamental change in the dynamical properties of the system on experimentally accessible timescales.
Our results indicate the presence of a crossover from a thermalizing regime to a prethermal pair-localized regime caused by positional disorder. 

The method for observing prethermal localization used in this work is inherently versatile and may also be applied to study thermalization in other systems. The signature that distinguishes thermalized from localized systems is the smooth dependence of the steady-state magnetization(, which is absent in the latter). This consideration becomes particularly crucial when the system's components, such as the pairs of spins in this study, experience rapid dephasing. This dephasing generally occurs on a much faster timescale compared to the build-up of entanglement between these components, resulting in thermalization. This insight calls for the reevaluation of claims made in~\cite{kucskoCriticalThermalizationDisordered2018, martinControllingLocalThermalization2022}, given that the relaxation of the magnetization in spatially disordered spin systems reflects only the dephasing but not the thermalization process. 

Notably, our system implements a critical case where the power law dependence of the interaction strength with distance $a$ equals the spatial dimension $d=3$. In this regime, theoretical results for large systems are scarce due to competing scales. In Appendix~\ref{appendix:vdW-data}, we show a similar experiment for a Van-der-Waals interacting system where $a=2d=6$. In this case, the magnetization behaves qualitatively as in the strongly disordered case of  $\alpha=3$ and also shows a sharp cusp. This indicates that prethermalization caused by localized pairs of spins is a robust effect independent of the spatial dimension as long as disorder is sufficiently strong. 

This study paves the way toward exploring the late-time dynamics of far-from-equilibrium systems with power-law interactions and disordered couplings, which are ubiquitous in nature. For these systems, it is yet an open question if they show (prethermal) many-body localization similar to the standard model of MBL where the on-site potential is disordered. 
Recent theoretical and numerical results indicate that localization and the consequent absence of thermalization are excluded in dimensions $d>1$ and for power law interactions~\cite{morningstarAvalanchesManybodyResonances2022, leonardSignaturesBathinducedQuantum2022}. However, the type of spatial disorder investigated in this study differs significantly from that in traditional MBL systems, rendering most conventional arguments about instability and eventual thermalization not directly applicable. Intriguingly, first numerical studies~\cite{braemerPairLocalizationDipolar2022} suggest that for the type of disorder studied here, localization effects are surprisingly robust to finite size drifts, a significant issue for the numerical investigation of the standard model of MBL. 
To draw parallels between our findings of prethermalization and prethermal MBL, it will be decisive to investigate the scaling of the relaxation timescale with the strength of disorder which is expected to be exponential in the case of prethermal MBL~\cite{longPhenomenologyPrethermalManyBody2022}. However, a proper definition of the strength of disorder in case the disordered couplings, opposed to disordered on-site detuning, remains an open question. 
Finally, an exciting avenue for future research is to explore the relation between the slow relaxation dynamics observed in this work and quantum spin glasses. In quantum spin glasses, the combination of frustration, low energies and disorder leads to exceptionally slow relaxation dynamics, a phenomenon  being highly relevant to the approach of quantum computation via quantum annealing~\cite{rajakQuantumAnnealingOverview2022, bernaschiQuantumTransitionTwoDimensional2023}. 

\clearpage  
\section*{Methods}\label{sec:methods}
Here we provide further details on the numeric simulations, the experimental protocol and the spatial configuration of the Rydberg cloud.

\textbf{Details on experimental implementation.} We start the experiment by trapping $10^6$ Rubidium-87 in a cigar shaped dipole trap with a diameter of \SI{300}{\micro\meter} (long axis) and \SI{70}{\micro\meter} (short axis) at a temperature of \SI{10}{\micro\kelvin}. 
We consider this gas to be frozen since the atoms move only a distance of $d_{\text{kin}} = t_{\text{exp}}\sqrt{\frac{3 k T}{m}} = \SI{0.5}{\micro\meter}$ during an experimental cycle of $t_{\text{exp}}=\SI{10}{\micro\second}$ which is small compared to the Rydberg blockade radius of $r_{\mathrm{bl}}\approx\SI{5}{\micro\meter}$. 
After optically pumping the atoms into the state $\ket{5S (F=2,m_F=2)}$, we optically excite the atoms to the spin state $\ket{\downarrow}$ via a two-photon off-resonant excitation process (single-photon detuning of $\SI{98}{\mega\hertz}$ and two-photon Rabi frequency of \SI{1}{\mega\hertz}). 
A global microwave $\pi/2$-pulse prepares the fully polarized initial state $\ket{\psi_0}=\ket{\rightarrow_x}^{\otimes N}$.  For the dipolar interacting spin system, we couple the states $\ket{48S}$ and $\ket{48P}$ resonantly with a single-photon transition at \SI{35}{\giga\hertz}. This frequency is generated by mixing a \SI{5}{\giga\hertz} signal of the Keysight M8190A AWG with an Anritsu MG3697C signal generator. In the case of Van-der-Waals interactions, the state $\ket{61S}$ is coupled resonantly to $\ket{62S}$ via a two-photon transition at a microwave frequency of \SI{16.546}{\giga\hertz} which can be directly generated with a Keysight M8190A arbitrary waveform generator (AWG). 

The same microwave setup is used to realize the spin locking field where a phase shift of 90 degrees needs to be added such that the field aligns with the spins. This allows us to implement the transverse field term, Eq.~\eqref{eq:H_ext}, with field strengths up to $\Omega/(2\pi) = \SI{10}{\mega\hertz}$. After a time evolution $t$, the $x$-magnetization is rotated tomographically onto the $z$-axis by applying a second $\pi/2$-pulse with various phases. Finally, the magnetization is obtained from a measurement of the population of one of the two spin states via field ionization, and the other spin state is optically deexcited to the ground state. A visual representation of the measurement protocol can be found in Fig.~\ref{fig:figure2}~\textbf{a}, and a more detailed explanation of the determination of the magnetization was reported in a previous publication~\cite{signoles2021}. 

\textbf{Details on the Rydberg distribution.} 
In this work, we can tune the disorder with the Rydberg blockade effect, which imposes a minimal distance $r_{\mathrm{bl}}$ between the spins. At small blockade radius, the spins are distributed randomly in the cloud, while a large radius introduces strong correlation between the atom positions and, hence, the coupling strength. To quantify the disorder strength, we compare the blockade radius to typical interparticle distance, which can be estimated from the Wigner-Seitz radius $a_0 = [3/(4\pi\rho)]^{1/3}$. We adjust this parameter in our experiment by controlling the Rydberg fraction, which is dependent on the excitation time $t_{\mathrm{exc}}$. In addition, we tune the Rydberg density $\rho$ by varying the volume of the ground state atoms with a short time-of-flight period after turning off the dipole trap and before exciting to the Rydberg states. We measure the resulting Rydberg density through depletion imaging~\cite{Ferreira-Cao2020DepletionImaging} where we deduce the Rydberg distribution from the missing ground state atoms after Rydberg excitation. The measured parameters of the Rydberg distribution are presented in detail in Table~\ref{tab:experimental_parameters} in the appendix.

To estimate the Rydberg blockade radius, we model the excitation dynamics by the simplified description introduced in \cite{signoles2021} which assumes a hard-sphere model for the Rydberg blockade effect. This model sets an upper limit on the blockade radius $r_{\mathrm{bl}}=\sqrt[6]{\frac{C_6}{\Gamma_{\mathrm{eff}}}}$ by estimating the effective linewidth of the laser, based on the duration of the excitation pulse and power broadening. The latter is calculated self-consistently, taking into account the enhancement factor induced by collective Rabi oscillations within a superatom~\cite{Weimer2008QuantumCritical, Garttner2012FinitesizeEffects}. 

This established model of the Rydberg cloud can be benchmarked using the experimentally measured time evolution, which is known to be well described by semiclassical Discrete Truncated Wigner Approximation (DTWA) in case no locking field is applied~\cite{signoles2021} (see Fig.~\ref{fig:comparison_dtwa} in appendix~\ref{app:semiclassical_simulations}). This simulation is highly sensitive with respect to the blockade radius and the density, and can therefore be used to determine these experimental parameters in case of weak and strong disorder. 
From the excitation model, we can also compute the median of the nearest neighbor interaction strength $J_{\mathrm{median}}$ which ranges from \SI{1.1}{\mega\hertz} to \SI{2.8}{\mega\hertz} depending on the experimental setting (see table~\ref{tab:experimental_parameters}). The resulting time evolution can be considered unitary for up to \SI{10}{\micro\second}, which is an order of magnitude larger than the timescale of the experiment $2\pi/J_\mathrm{median}$. 

\clearpage
\appendix

\begin{table*}
\resizebox{\textwidth}{!}{%
\begin{tabular}{lccc}
\toprule
{} &                                             Dipolar interactions (weak disorder) &                                            Dipolar interactions (strong disorder) &                                                       Van-der-Waals interactions \\
\midrule
Rydberg states        &                                $\ket{48S_{1/2}} \leftrightarrow \ket{48P_{3/2}}$ &                                 $\ket{48S_{1/2}} \leftrightarrow \ket{48P_{3/2}}$ &                                $\ket{61S_{1/2}} \leftrightarrow \ket{62S_{1/2}}$ \\
decay rate $\Gamma/(2\pi)$   &                                                          \SI{0.018}{\mega\hertz} &                                                           \SI{0.018}{\mega\hertz} &                                                         \SI{0.0096}{\mega\hertz} \\
$t_{\mathrm{exc}}$    &                                                         $\SI{10}{\micro\second}$ &                                                           $\SI{1}{\micro\second}$ &                                                          $\SI{5}{\micro\second}$ \\
Excitation volume     &  $\SI{59}{\micro\meter}\times \SI{44}{\micro\meter}\times \SI{36}{\micro\meter}$ &  $\SI{59}{\micro\meter}\times \SI{34}{\micro\meter} \times \SI{30}{\micro\meter}$ &  $\SI{69}{\micro\meter}\times \SI{43}{\micro\meter}\times \SI{37}{\micro\meter}$ \\
$N_{\mathrm{Ryd}}$    &                                                                             6895 &                                                                               775 &                                                                             2907 \\
$r_{\mathrm{bl}}$     &                                                       $\SI{ 4.6 }{\micro\meter}$ &                                                        $\SI{ 5.0 }{\micro\meter}$ &                                                       $\SI{ 5.7 }{\micro\meter}$ \\
$a^0$                 &                                                      $\SI{ 6.8 } {\micro\meter}$ &                                                      $\SI{ 11.2 } {\micro\meter}$ &                                                      $\SI{ 7.8 } {\micro\meter}$ \\
$J_{\mathrm{median}}/(2\pi)$ &                                                       $\SI{ 2.8 }{\mega\hertz}$ &                                                         $\SI{ 1.1 }{\mega\hertz}$ &                                                        $\SI{ 0.5 }{\mega\hertz}$ \\
\bottomrule
\end{tabular}

}
\caption{Experimental parameters. $t_{\mathrm{exc}}$ specifies the duration of the optical excitation to the Rydberg state, the Rydberg volume is specified by the radii ($1/e^2$) of the Rydberg cloud, $N_{\text{Ryd}}$ denotes the derived Rydberg number, $r_{\mathrm{bl}}$ the blockade radius and $J_{\text{median}}$ the obtained median nearest-neighbor interaction.}
\label{tab:experimental_parameters}
\end{table*}

\section{Semiclassical DTWA simulations}
\label{app:semiclassical_simulations}

\begin{figure*}[!htb]
    \centering
    \includegraphics[scale=0.9]{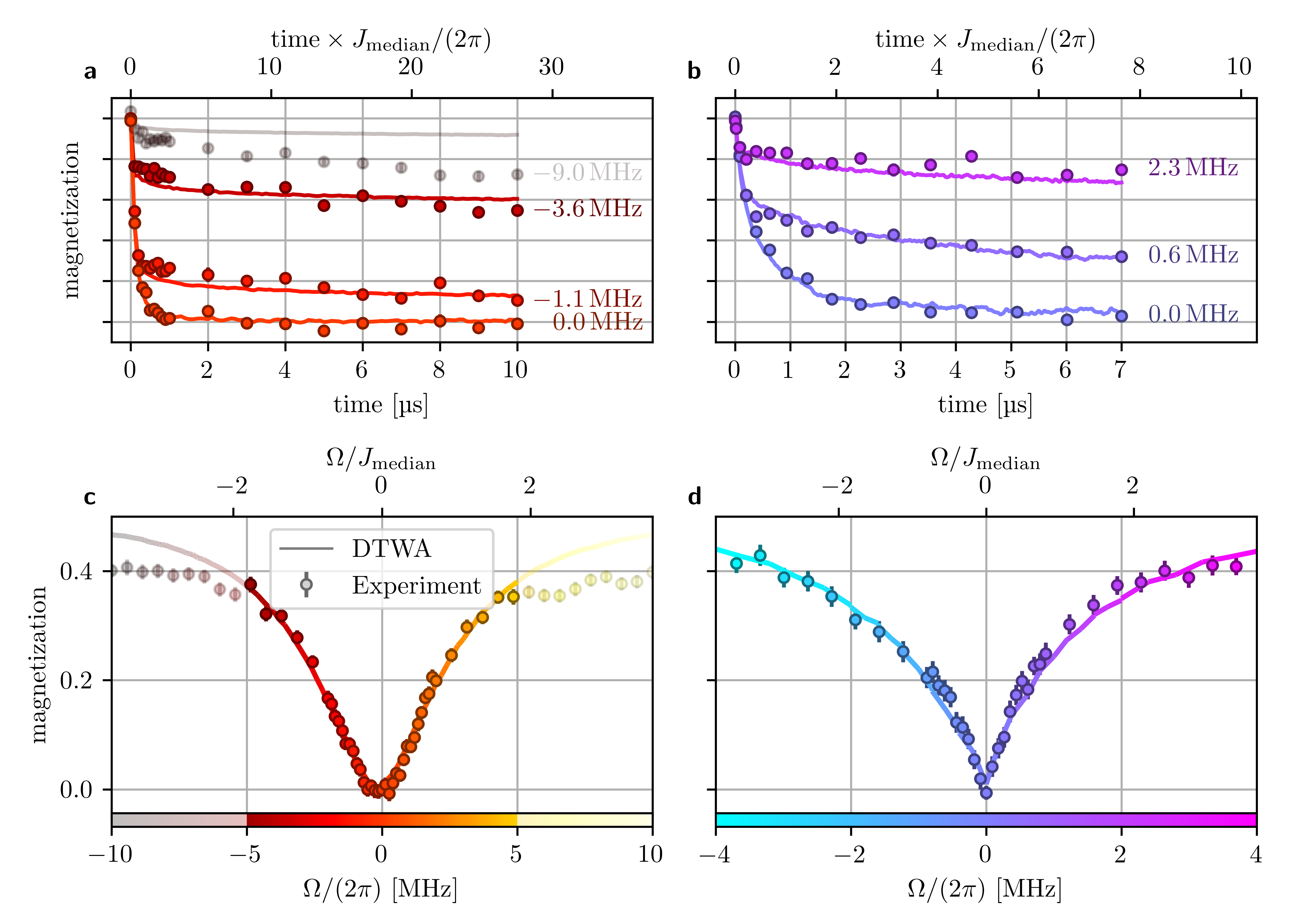}
    \caption{\textbf{Simulation of the experimental data shown in the main text with DTWA simulations} Time evolution of the magnetization in case of weak (\textbf{a}) and strong (\textbf{b}) disorder. The dependence of the late-time magnetization are shown in \textbf{c} (weak disorder) and \textbf{a} (strong disorder). The experimental parameters are shown in Table~\ref{tab:experimental_parameters}.}
    \label{fig:comparison_dtwa}
\end{figure*}

In previous work~\cite{Orioli2018RelaxationIsolateda, signoles2021}, we could show that the semiclassical Discrete Truncated Wigner Approximation (DTWA) is well suited to describe the relaxation of the magnetization under the interaction Hamiltonian \eqref{eq:H_int} defined in the main text. The main principle of DTWA is to sample classical time evolutions over different initial states such that the quantum uncertainty of the initial state is respected~\cite{Schachenmayer2015ManyBodyQuantum}.  In Fig.~\ref{fig:comparison_dtwa}, we compare the time evolution obtained from DTWA simulations to the experimental data (red dots) in the case of weak (left panels) and strong disorder (right panels). It turns out that the resulting dynamics depend sensitively on the blockade radius and on density. However, the same fitted parameters describes the time evolution for different locking fields (top panels) and the dependence of the late-time magnetization on the field strength. As mentioned in the main text, the observed discrepency of DTWA simulations and experimental data observed for large fields in the weakly disordered regime (Fig.~\ref{fig:comparison_dtwa}~\textbf{c}) can be most likely attributed to experimental imperfections such as coupling to other Rydberg states due to power broadening of the microwave transition.

\section{Data for Van der Waals interactions}
\label{appendix:vdW-data}

By encoding the spin degree of freedom in different Rydberg states, it is possible to realize different Hamiltonians with different range of interactions. In addition to a dipolar interacting Hamiltonian with $a=d=3$ as shown in the main text, we can also create a spin system with less long-range Van der Waals interactions. For this purpose, we couple the Rydberg state $\ket{\downarrow} = \ket{61S}$ to $\ket{\uparrow} = \ket{62S}$ which results in a Heisenberg XXZ Hamiltonian as described by Eq.~\eqref{eq:H_int} with  $\delta = \num{-0.7}$, $a=6$ and $C_6/(2\pi) = \SI{507}{\giga\hertz \micro\meter\tothe{6}}$ (see also Table~\ref{tab:experimental_parameters} for an overview over the experimental parameters).

\begin{figure*}[!htb]
    \centering
    \includegraphics[scale=1]{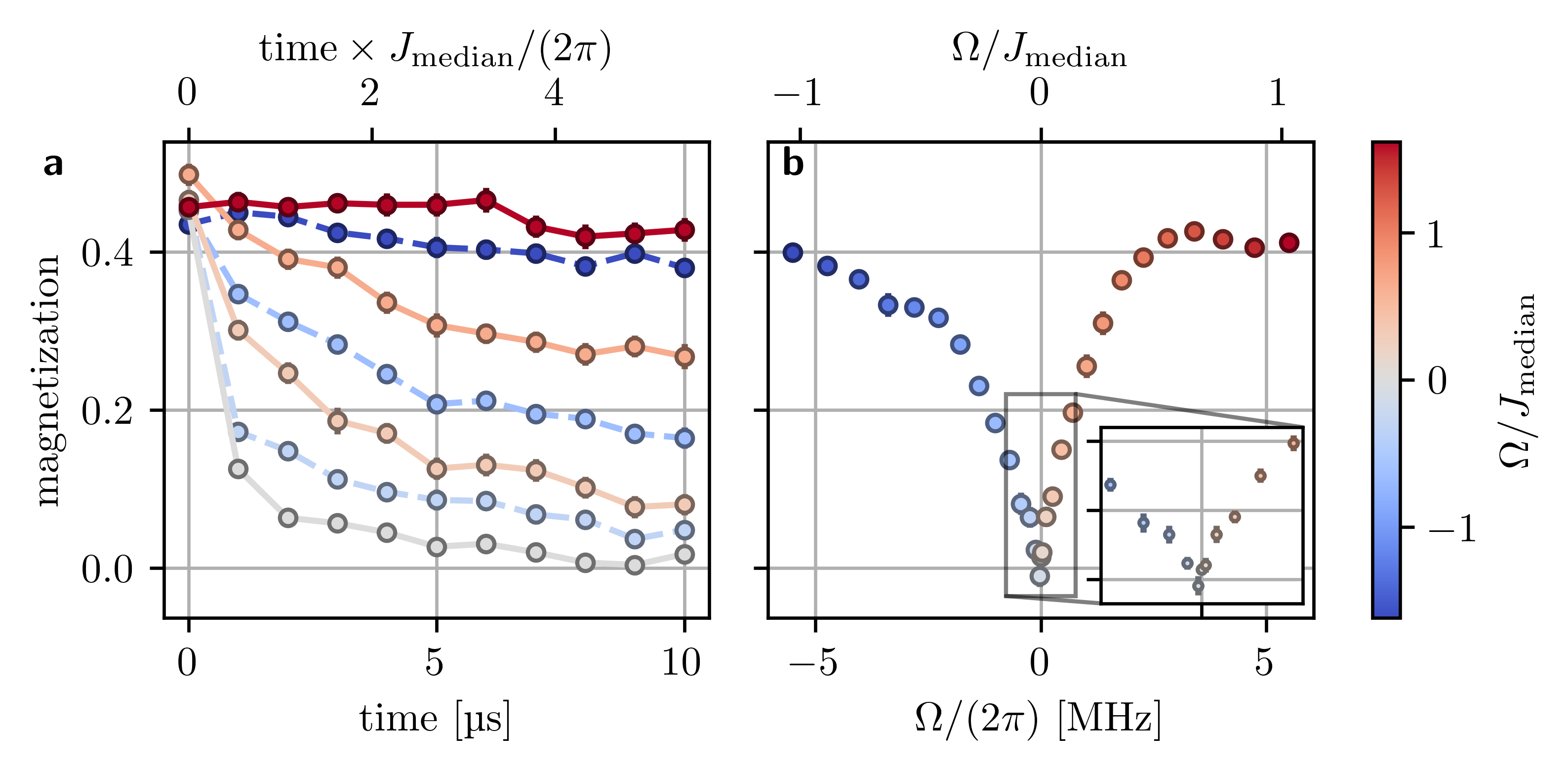}
    \caption{\textbf{Experimental data for a quantum spin system with Van der Waals interactions. a,} Measured spin relaxation dynamics for varying transverse field strengths ranging from $\Omega/(2\pi)=\SI{-5.5}{\mega\hertz}$ (dark blue) to \SI{5.5}{\mega\hertz} (dark red).
    \textbf{b,} Magnetization after \SI{10}{\micro\second} as a function of field strength $\Omega$ (see Table~\ref{tab:experimental_parameters} for a summary of the experimental parameters). The inset shows a zoom into the data for small values of $\Omega$. }
    \label{fig:van-der-waals}
\end{figure*}

Similar to the experimental results presented in the main article, also the Van der Waals interacting system shows a slow relaxation dynamics on a timescale of $\approx 2\pi/J_{\mathrm{median}}$ (see Fig.~\ref{fig:van-der-waals}~\textbf{a}). Applying an external field $\Omega$ also slows down the relaxation dynamics considerably. It should be noted, that the external field has to be realized by a two-photon microwave transition as the transition between the two spin states is dipole forbidden. Therefore, the single photon Rabi frequencies are required to be much larger compared to the dipolar interacting spin system, which might potentially lead to a stronger coupling to different Rydberg states inducing addition decay of the magnetization, especially at late times. 

The dependence of the late-time magnetization (taken after \SI{10}{\micro\second}) on the spin locking field $\Omega$ of the Van der Waals interacting system is shown in Fig.~\ref{fig:van-der-waals}~\textbf{b}. Compared to the dipolar interacting case presented in the main text, the curve is even more asymmetric. This effect can be explained by the isotropic repulsive interactions in the Van der Waals case, whereas dipolar couplings vary as $1-3\cos(\theta)^2$ depending on the angle $\theta$ between the inter-spin axis and the quantization axis. Most importantly, also the Van der Waals interacting system features a sharp cusp around $\Omega=\SI{0}{\mega\hertz}$. In this regard, the curve strongly resembles the case of strong disorder in dipolar interacting systems presented in the main text. At first sight, this result might be surprising as the spin system is even more blockaded with a ratio of blockade radius to typical interaction range of $r_{\mathrm{bl}}/a_0 = 5.7/7.8 = 0.73$ than the weakly disordered dipolar system where $r_{\mathrm{bl}}/a_0 = 4.6/6.8 = 0.68$\footnote{Due to the Van der Waals interactions being a second order process, the typical interaction strength are much weaker compared to the dipolar interacting case. To compensate for this effect, we increase the density which increases the interaction strength.}.  However, the shorter-range interaction increases the effective disorder in the system as the nearest-neighbor interaction becomes much stronger compared to the next-nearest neighbor coupling. This proves that, especially for short-range interactions decaying faster than $a=d$, the existence of a prethermal state is a ubiquitous phenomenon in spatially disordered quantum spin systems.

\section{Derivation of the effective model}
\label{app:pair_model}

In this appendix, we derive how to describe the system in terms of localized pairs, which constitute the approximate local integrals of motion for the GGE description of the system. Starting from the physics of a single, isolated pair, we will derive the GGE, the description in terms of thermal pairs, and the self-consistent mean-field equations, which partly take into account interactions beyond the nearest neighbor. 
This approximation provides an intuitive picture that allows us to explain all the observed features of the long-time magnetization (positivity, cusp, asymmetry).

For a single interacting pair, in the basis $\{\ket{\rightarrow\rightarrow},\ket{\rightarrow\leftarrow},\ket{\leftarrow\rightarrow},\ket{\leftarrow\leftarrow}\}$, Hamiltonian~\eqref{eq:H_int} reads
\begin{align}
    \hat{H}_{\mathrm{pair}} &= 4 J\left(\Delta \hat{s}_x^{(1)}\hat{s}_x^{(2)} + \hat{s}_y^{(1)}\hat{s}_y^{(2)} + \hat{s}_z^{(1)}\hat{s}_z^{(2)}\right) + \Omega \sum_{i=1}^2\hat{s}_z^{(i)} \\
    &= \begin{pmatrix}
        J + \Omega & 0 & 0 & J(\Delta - 1)\\
        0 & -J & J(\Delta + 1) & 0 \\
        0 & J(\Delta + 1) & -J & 0 \\
        J(\Delta - 1) & 0 & 0 & J - \Omega
    \end{pmatrix}
\end{align}
where we defined $J = J_{12}/4$.
Out of the four eigenstates of this Hamiltonian, only two have non-zero overlap with the initial state $\ket{\rightarrow\rightarrow}$ (see table~\ref{tab:pair_eigenstates}). Therefore, each interacting pair can be seen as an effective two-level system on its own, with a modified interaction between these "renormalized" spins. This ansatz of diagonalizing the strongest interacting pairs first can be seen as a first step in a real-space strong-disorder renormalization group treatment~\cite{pekkerHilbertGlassTransitionNew2014, vasseurParticleholeSymmetryManybody2016a, vasseurQuantumCriticalityHot2015, voskManyBodyLocalizationOne2013}. Here, we do not aim to proceed further in this renormalization scheme, but instead, we use the basis of eigenstates of strongly interacting pairs to derive an intuitive understanding of the physics within mean-field theory. 

\begin{table*}[t]
  \centering
  \begin{tabular}{c|c|c|c}
    Eigenvalue $E_i$ & Eigenvector $\ket{\psi_i}$ & Occupation $|\braket{\psi_0}{\psi_i}|^2$ & Magnetization $\expval{\hat{S}_x}_{\psi_i}$ \Bstrut\\ 
    \hline\Tstrut\Bstrut
    
    $J\Delta$ & $\dfrac{1\Tstrut}{\sqrt{2}\Bstrut} \left( \ket{\rightarrow\leftarrow} + \ket{\leftarrow\rightarrow}\right)$ &0 & 0\\
    
    $-J(2 + \Delta)$ & $\dfrac{1\Tstrut}{\sqrt{2}\Bstrut} \left( \ket{\rightarrow\leftarrow} - \ket{\leftarrow\rightarrow}\right)$  & 0 & 0 \\
    
    $J-\sqrt{\Omega^2+J^2\left(\Delta-1\right)^2}$ & $\sqrt{\dfrac{1}{2} - \dfrac{\Omega}{2\sqrt{\Omega^2 + j^2}}}\ket{\rightarrow\rightarrow} + \sqrt{\dfrac{1}{2} + \dfrac{\Omega}{2\sqrt{\Omega^2 + j^2}}}\Tstrut\Bstrut\ket{\leftarrow\leftarrow}$  & $\dfrac{1}{2} - \dfrac{\Omega}{2\sqrt{\Omega^2 + j^2}}$ & $-\dfrac{\Omega}{2 \sqrt{\Omega^2+j^2}}$ \\
    
    $J+\sqrt{\Omega^2+J^2\left(\Delta-1\right)^2}$ & $\sqrt{\dfrac{1}{2} + \dfrac{\Omega}{2\sqrt{\Omega^2 + j^2}}}\ket{\rightarrow\rightarrow} + \sqrt{\dfrac{1}{2} - \dfrac{\Omega}{2\sqrt{\Omega^2 + j^2}}}\Tstrut\Bstrut\ket{\leftarrow\leftarrow}$  & $\dfrac{1}{2} + \dfrac{\Omega}{2\sqrt{\Omega^2 + j^2}}$ & $\dfrac{\Omega}{2 \sqrt{\Omega^2+j^2}}$
  \end{tabular}
  \caption{Properties of the four eigenstates of a single interacting spin pair. To simplify notation, we introduced $j = J\left(\Delta - 1\right)$.}
  \label{tab:pair_eigenstates}
\end{table*}

\begin{figure}
\centering
\includegraphics[width=1\linewidth]{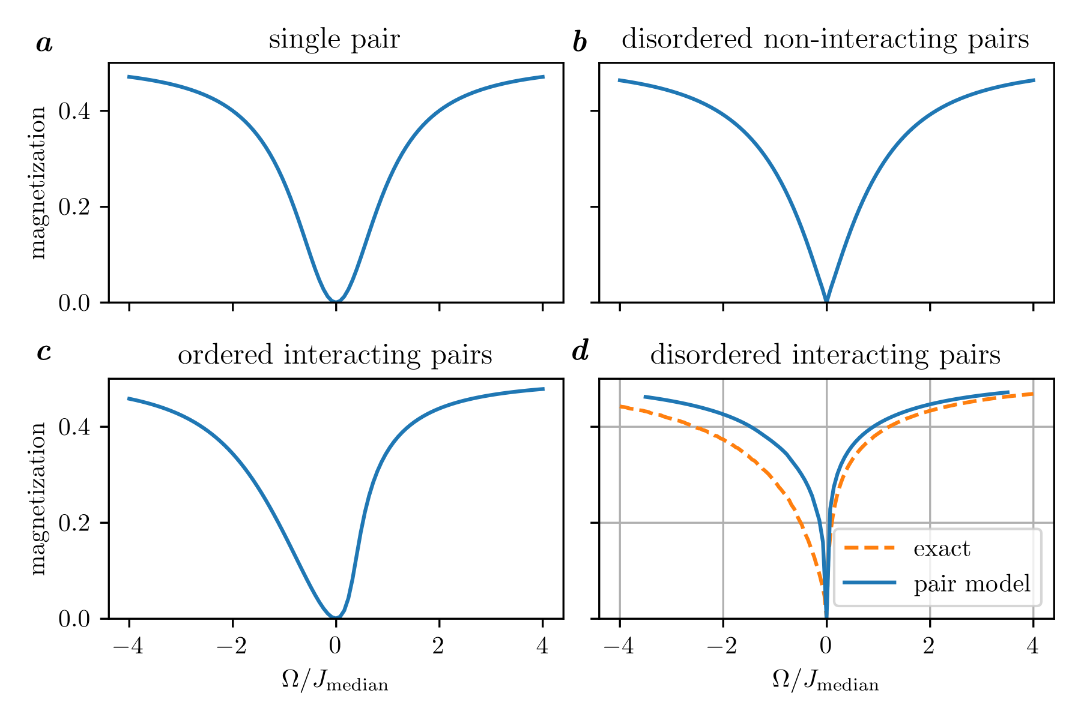}
\caption{
	The diagonal ensemble expectation value of the magnetization as a function of applied external field $\Omega$ for a single pair (\textbf{a}), a disorder average of single pairs with interaction chosen randomly in the interval $J \in [0, 1]$ (\textbf{b}), a system of identical pairs that interact with mean field interaction $J_{\mathrm{inter}} = 1.5*J$ (\textbf{c}), and a realistic random distribution with power-law interactions, as described in the text (\textbf{d}). For the latter, the dashed orange line shows the full quantum mechanical solution obtained by exact diagonalization for the same system.
}
\label{fig:cusp_toy_model}
\end{figure}

\textbf{Diagonal ensemble:} In contrast to a single spin which does not show any dynamics, a strongly interacting pair features oscillatory dynamics. Using the definition given in the main text, we can calculate the diagonal ensemble expectation value for single pair:
\begin{equation}
    \overline{\langle \hat{S}_x \rangle_{\mathrm{pair}}} = \frac{\Omega^2}{2(\Omega^2 + j^2)}
    \label{eq:diagonal_pair}
\end{equation}
where we introduced $j = J\left(\Delta - 1\right)$.
It should be noted that this diagonal ensemble does not describe the steady-state but rather the time average over the oscillations. The magnetization expectation value predicted by the diagonal ensemble of a single interacting pair represents an inverted Lorentz profile with width $j/2$, which features a quadratic dependence on $\Omega$ around zero (see Figure~\ref{fig:cusp_toy_model}~\textbf{a}). However, if we average over multiple pairs with different interaction strengths $j$, the diagonal ensemble value becomes more meaningful since we can assume that the different oscillation frequencies dephase. Also, the behavior of the magnetization changes: For example, assuming a uniform distribution of $j \in [0, \Delta_j]$\footnote{For distributions like $j \in [j_{\mathrm{min}}, \Delta_j]$ that do not feature arbitrary small interaction strengths, a small region of approximate size $\Omega < |\dfrac{j_{\mathrm{min}}}{\Delta_j}|$ exists where magnetization is a smooth function of external field.}, we obtain 
\begin{equation}
    \frac{1}{\Delta_j} \int_0^{\Delta_j} \overline{\langle \hat{S}_x \rangle_{\mathrm{pair}}} \, \mathrm{d}j = \frac{\Omega}{2\Delta_j} \arctan(\frac{\Delta_j}{\Omega})
\end{equation}
which shows the non-analytic cusp feature at $\Omega=0$ (see Figure~\ref{fig:cusp_toy_model}~\textbf{b}). Close to the non-analytic point, the magnetization increases linearly with a slope $\dfrac{\pi}{4 \Delta_j}$ inversely proportional to the width of the distribution of interaction strengths. Therefore, we can conclude that the non-analyticity is a direct consequence of disorder and the resulting broad distribution of nearest neighbor interaction strengths.

\textbf{Canonical and generalized Gibbs ensemble:} To calculate the properties of a system in thermal equilibrium, we evaluate the density matrix $\hat{\rho}_{\mathrm{canonical}}$ of the canonical ensemble
\begin{equation}
    \hat{\rho}_{\mathrm{canonical}} = \frac{\sum_i e^{-\beta E_i} \ket{\psi_i}\bra{\psi_i}}{ \sum_i e^{-\beta E_i}}
    \label{eq:canonical_ensemble}
\end{equation}
where $\beta$ is the inverse temperature of the system. For a single pair of spins, this ensemble can be used to calculate the expectation value of the magnetization: 
\begin{equation}
   \langle\hat{s}_x^p\rangle_{\text{canonical}}(\beta) = - \frac{h}{2\sqrt{h^2 + j^2}} \tanh\left(\sqrt{h^2 + j^2} \beta\right)
\end{equation}
In a system coupled to a thermal bath, the inverse temperature $\beta$ would be determined by the temperature of the bath. However, in a closed the system, the energy is conserved, which fixes the inverse temperature such that the energy of the canonical ensemble equals the energy of the initial state. In a generalized Gibbs ensemble, where the energy of each pair of spins is conserved, this leads to the equation 
\begin{align}
    \langle\hat{H}_{\mathrm{pair}}\rangle_{\text{canonical}}(\beta) \overset{!}{=}  \langle\hat{H}_{\mathrm{pair}}\rangle_{\ket{\psi_0}} \label{eq:canonical_condition}\\ 
    \Leftrightarrow - \sqrt{h^2 + j^2}\tanh\left( \sqrt{h^2 + j^2}\beta \right) + J \overset{!}{=}  h + J
\end{align}
This equation can be solved analytically and results in exactly the diagonal ensemble from Eq.~\ref{eq:diagonal_pair}. This result is not surprising considering the following argument: Only two out of four eigenstates of the pair of spins can be occupied due to symmetry arguments. Thus, any mixture of these states is completely determined by only two variables. Out of those, one is fixed by normalization and the other by energy, and all ensembles are strictly equivalent. 

In the generalized Gibbs ensemble, we have considered an ensemble of perfectly isolated pairs, where each pair $i$ has equilibrated to a different inverse temperature $\beta_i$. A first approximation to estimate the magnetization of a thermalized ensemble of disordered spins can be obtained by assuming weak interactions between each pair of spins that do not affect the eigenstates but lead to thermalization such that every spin relaxes to a canonical ensemble with one global $\beta = \beta_i$ for all pairs $i$. In this case, eq.~\eqref{eq:canonical_condition} has to be solved for $\beta$ for the sum of all pairs: 
\begin{equation}
    \sum_i \langle\hat{H}_{\mathrm{pair, i}}\rangle_{\text{canonical}}(\beta) \overset{!}{=}  \sum_i \langle\hat{H}_{\mathrm{pair, i}}\rangle_{\ket{\psi_0}}  .
\end{equation}
For this value of $\beta$, the canonical ensemble expectation value for the average magnetization can be calculated using equation \eqref{eq:canonical_ensemble}. 

\textbf{Self-consistent mean-field equations:} To obtain an even more realistic model and to understand additional features like the asymmetry of the cusp, we add a mean-field interaction between pairs. For this purpose, we replace the external field with an effective mean-field acting on spin $i$:
\begin{equation}
    \Omega \rightarrow \Omega_i = \Omega + \sum_j J_{ij}^{\mathrm{inter}} \langle \hat{s}_x^{(j)} \rangle
    \label{eq:mean-field}
\end{equation}
As a first example, we may consider a periodic chain of equally spaced pairs where all pairs are identical and the mean-field shift arising from interactions between the pairs is $J^{\mathrm{inter}}$. In this case, the diagonal ensemble expectation value can be calculated by solving the self-consistent equation
\begin{equation}
    \overline{\langle \hat{S}_x \rangle} = \frac{1}{2}\frac{\left(\Omega + J^{\mathrm{inter}} \overline{\langle \hat{S}_x \rangle}\right)^2}{\left(\Omega + J^{\mathrm{inter}} \overline{\langle \hat{S}_x \rangle}\right)^2 + j^2}.
\end{equation}
Since the right-hand side of the equation only contains squares, the magnetization is still positive or zero. Therefore, for positive external fields $\Omega$, the effective field is larger than the external field ($\Omega_i \ge \Omega$), leading to an enhanced spin locking effect. Consequently, mean-field leads to an increased magnetization compared to the case of independent pairs. For negative $\Omega$, the external field is anti-aligned with the mean-field, and the resulting magnetization is decreased. Thus, the dependence of the magnetization as a function of field strength is asymmetric (see Figure~\ref{fig:cusp_toy_model}~\textbf{c}. In conclusion, we can attribute the asymmetry to mean-field interaction between different pairs.

In order to model the disordered spin system realized experimentally, we apply the pair model to an ensemble of spins with randomly chosen positions. We cluster the spins $i$ into pairs $p$ in such a way that the sum over all pair distances is minimized. Naturally, the interaction $j_p$ of a pair $p$ consisting of spins $i$ and $j$ is given by the interaction strength between the spins. The interaction strength $J_{pq}^{\mathrm{inter}}$ between pair $p$ and $q$ can be obtained from the strongest interaction $J_{ij}$ where spin $i$ is in pair $p$ and $j$ in $q$ respectively. Now, we solve the system of self-consistent equations
\begin{equation}
    \overline{\langle \hat{s}_x^{p} \rangle} = \frac{1}{2}\frac{\left(\Omega + \sum_q (J_{pq}^{\mathrm{inter}} \overline{\langle \hat{s}_x^{q} \rangle })\right)^2}{\left(\Omega + \sum_q (J_{pq}^{\mathrm{inter}} \overline{\langle \hat{s}_x^{q} \rangle })\right)^2 + j_{p}^2}.
\end{equation}
The resulting magnetization curve obtained after disorder averaging (see blue line in Figure~\ref{fig:cusp_toy_model}~\textbf{d} closely resembles the exact diagonal ensemble prediction (orange line). Importantly, all qualitative features are captured, including a positive magnetization which is asymmetric with respect to the external field and shows a sharp cusp at zero field. 
The remaining discrepancy between the pair model and the exact solution, in particular the stronger asymmetry of the exact solution, can be attributed to clusters of spins containing more than two atoms where quantum fluctuations decrease the magnetization even further than predicted by the pair mean-field model.

\newpage
\bibliography{cusp}

\end{document}